\documentclass[12pt]{article}
\usepackage{amsmath}
\usepackage{amsthm}
\usepackage{amssymb}
\usepackage{mathrsfs}
\usepackage{authblk}
\usepackage{cite}
\usepackage{pdfsync}
\usepackage{enumitem}
\setlist{parsep=0pt,itemindent=0pt}

\usepackage{graphicx}

\theoremstyle{definition}

\numberwithin{equation}{section}
\numberwithin{thm}{section}
\numberwithin{lemma}{section}
\numberwithin{prop}{section}
\numberwithin{cor}{section}
\numberwithin{rmk}{section}
\numberwithin{defn}{section}
\numberwithin{exa}{section}

\newcommand{\gen}[1]{\partial_{#1}}
\newcommand{\curl}[1]{ \left\{#1\right\} }
\newcommand{\lie}{\mathfrak g}

\DeclareMathOperator{\SU}{SU}
\DeclareMathOperator{\su}{su}

\DeclareMathOperator{\sn}{sn}

\begin{document}
\pagenumbering{arabic}
\clearpage
\thispagestyle{empty}

\title{Explicit solutions with non-trivial phase of the inhomogeneous coupled two-component NLS system}

\author[1]{J.~Belmonte-Beitia\thanks{Juan.Belmonte@uclm.es}}
\author[2]{F.~G\"ung\"or\thanks{gungorf@itu.edu.tr}}
\author[3]{P.J. Torres\thanks{ptorres@ugr.es}}

\affil[1]{Departmento de Matem\'aticas, E.T.S. Ingenier\'ia Industrial and Instituto de Matem\'atica Aplicada a la Ciencia y la Ingenier\'ia (IMACI), Universidad de Castilla-La Mancha, Spain}
\affil[2]{Department of Mathematics, Faculty of Science and
Letters, Istanbul Technical University, 34469 Istanbul, Turkey}
\affil[3]{Departamento de Matem\'atica Aplicada and Research Unit ``Modeling Nature'' (MNat), Universidad de
Granada, 18071 Granada, Spain}

\date{\today}

\maketitle

\begin{abstract}
In this article, we construct novel explicit solutions for nonlinear Schr\"odinger systems with spatially inhomogeneous nonlinearity by means of the Lie symmetry method. We focus the attention to solutions with non-trivial phase, which have been scarcely considered in the related literature. To get started, the theoretical method based on Lie symmetries is exposed, thus reducing the problem to the integrability of an ODE. The non-trivial phase introduces a singular term into the ODE. Then, the method is used to construct new families of analytical solutions. Some illustrative examples are provided.
\end{abstract}

\vspace{1,5mm}




\section{Introduction}

The nonlinear Schr\"odinger equation (NLS) is one of the most important equations of mathematical physics as a model arising in many scenarios\cite{sulem}. This equation appears in the modeling of many physical phenomena with applications to different fields, such as semiconductor physics \cite{Brezzi}, nonlinear optics \cite{Kivshar}, condensation of Bose--Einstein \cite{Dalfovo}, quantum mechanics \cite{Rosales}, plasma physics \cite{Dodd} or  gravitation theory \cite{belinski_verdaguer_2001}, to cite just a few examples.

In the latter years,  an enormous interest on the study of NLS equations with nonlinear coefficients depending either on space, time or both has attracted the attention of many researchers. The main motivation comes from the applications of the model to the fields of Bose--Einstein condensates (BECs). In this field of application, the so-called Feschbach resonance management \cite{Inouye} allows for a precise control of the atomic interactions responsible for the strength of nonlinearities. This has been experimentally implemented leading to time dependent nonlinear coefficients and would allow for the generation of space or space–time dependent nonlinearities since the diffusion length can be continuously changed by varying the external or optical magnetic field. Moreover, the consideration of inhomogeneous nonlinearities has led to the prediction of many remarkable nonlinear phenomena in the last few years, either for time dependent \cite{Saito,Konotop,Pacciani,Kevrekidis_2005} or space dependent nonlinear coefficients \cite{Garnier,Malomed,Carpentier,Nuestro2}.

Although the model equations of interest are not integrable in general, there have been efforts to construct families of exact solutions of the NLS with spatially inhomogeneous coefficients. Specifically, group-theoretical methods based on Lie symmetries have been used to obtain solutions of NLS type equations in Refs. \cite{Nuestro2,Tang,Nuestro3,Hao}. However, up to now the applications of the technique have been restricted  to get solutions with trivial phase (i.e. $\theta\equiv 0$). Thus, the purpose of this paper is to extend the method and construct exact solutions for multicomponent systems with non-trivial phase ($\theta\neq 0$).

In BEC applications, these systems have enormous physical interest, since the model describes two hyperfine levels of an atomic BEC. Different papers have studied exact soliton solutions for similar systems in the case of homogeneous nonlinearities \cite{Kaup,Victor,BELMONTEBEITIA,Deconinck_2003,Matsuno,Nath}. In \cite{Belmonte-BeitiaPerez-GarciaBrazhnyi2011}, the authors report the existence of different analytical solutions as bright-bright, dark-dark and dark-bright solutions of the two-component system with spatially inhomogeneous nonlinearities. Very recently, in \cite{Rogue} vector rogue wave (RW) formation and their dynamics in Rabi coupled two-and three-species BECs with spatially varying dispersion and nonlinearity were studied. Also, in  \cite{Kartashov}, the authors considered the existence of two-component solitons in a medium with a periodic nonlinear coefficient.

Here we will go beyond previous studies calculating different analytical solutions for multicomponent systems with inhomogeneous nonlinearities and non-trivial phase. Some of these solutions correspond to dark-dark soliton and multi-peak bright soliton solutions. To our best knowledge, this is the first time that these sort of solutions are reported with non-trivial phase. As we said previously, our focus will be to construct solutions for applications to matter waves in BECs, but we think that our ideas can also be applied to nonlinear optics \cite{Kivshar}.

The paper is organized as follows. First, in Section 2, we introduce our physical model: a system of two coupled nonlinear Schr\"odinger equations with spatially inhomogeneous nonlinearities. Moreover, we construct the general solutions with non-trivial phase.  In Section 3, the general theory of the Lie symmetry analysis for our model problem is introduced.  Section 4 shows a particular case obtaining a reduction of the previous system to an integrable singular ordinary differential equations. Next, Section 5, discusses the particular case where $h_{ij}=1$, for $i,j=1,2$. In Section 6, we show some analytical solutions calculated using the results obtained in the previous sections which are of physical interest. Finally, in Section 7, we summarize our conclusions.

\section{The model}

We consider the one-dimensional system of coupled cubic Schr\"odinger equations
\begin{equation}\label{sys}
  \begin{split}
      & i \frac{\partial\psi_1}{\partial t}+\frac{\partial^2 \psi_1}{\partial x^2}=V_1(x)\psi_1+(g_{11}(x)|\psi_1|^2+g_{12}(x)|\psi_2|^2)\psi_1,\\
       & i \frac{\partial\psi_2}{\partial t}+\frac{\partial^2 \psi_2}{\partial x^2}=V_2(x)\psi_2+(g_{21}(x)|\psi_1|^2+g_{22}(x)|\psi_2|^2)\psi_2,
   \end{split}
\end{equation}
where $V_j(x)$ and $g_{ij}(x)$ are arbitrary real functions of $x$ and $\psi_1, \psi_2(t,x)\in \mathbb{C}$ are complex-valued functions of $(t,x)\in \mathbb{R}\times\mathbb{R}$.

The special case when $V_i(x)=0$ and $g_{11}=g_{12}=1$  and $g_{12}=g_{21}=\sigma$ (constant), which from now on will be referred to standard coupled NLS (cNLS) system,
\begin{equation}\label{sigma-sys}
  i\psi_{1,t}+\psi_{1,xx}=(\sigma|\psi_1|^2+|\psi_2|^2)\psi_1, \quad i\psi_{2,t}+\psi_{2,xx}=(|\psi_1|^2+\sigma|\psi_2|^2)\psi_2
\end{equation}
is of particular significance in applications. For $\sigma=0$ the system reduces to two uncoupled integrable NLS equations. This system with $\sigma=1$ was proved to be integrable by Manakov \cite{Manakov1974}. The corresponding  system is known as Manakov system and can be written in vector notation as
\begin{equation}\label{vNLS}
\quad i\psi_t+\psi_{xx}=({\psi}^{\dag }\psi)\psi,
\end{equation}
where the dagger denotes hermitian conjugation. $\psi\in\mathbb{C}^2$ is a two-component complex vector.

We introduce the moduli $R_j(t,x)$ and phases $\phi_j(t,x)$ of the components of  $\psi$ by putting
\begin{equation}\label{mod-phase}
 \psi_{j} = R_{j} e ^{ i \phi_ j } , \quad 0 \leq R_{j} < \infty , \quad 0 \leq \phi_{j} \leq 2 \pi, \quad j=1,2.
\end{equation}
System \eqref{sys} for arbitrary coefficients is invariant under the abelian Lie point symmetry algebra $\lie_0$ generated by the vector fields
\begin{equation}\label{symalg}
  T=\gen t, \quad M_j=\gen {\phi_j},  \quad j=1,2.
\end{equation}
When it happens $V_1(x)=V_2(x)$, $g_{ij}(x)=g(x)$ for some $g(x)$, the symmetry algebra $\lie_0$ gets enlarged by the gauge  group $\SU(2)$, realized by the nonabelian constant gauge transformations
\begin{equation}\label{su2}
  \widetilde { \psi } = U \psi, \quad U ^ {\dag } U = I , \quad U \in \mathbb { C } ^ { 2 \times 2 },
\end{equation}
where $U$ is a constant unitary matrix. Then, we have the direct sum algebra $\lie=\lie_0\oplus \su(2)$ as the symmetry algebra.

The general system with inhomogeneous coefficients \eqref{sys} is not integrable in general, and many papers have been devoted to the search of exact solutions with physical meaning. A standard procedure is to look for translationally invariant solutions under the one-parameter subgroup $\exp\curl{\varepsilon(T+\mu_1M_1+\mu_2M_2)}$ generating time translations combined with changes of phase: $t\to t+\varepsilon$, $\phi_j\to \phi_j+\varepsilon \mu_j$. They are solutions (uniformly propagating coherent wave structures \cite{LiuTorresXing2018}) of the form
\begin{equation}\label{uni-wave}
  \psi_j(t,x)=R_j(x)  \exp \left[ i \left( \theta _ { j } ( x ) + \mu _ { j } t \right) \right],  \quad j = 1,2.
\end{equation}
Without mentioning any invariance property of solutions, this special ansatz has been used in a number of papers (See for example \cite{PorubovParker1999, ChristiansenEilbeckEnolskiiKostov2000}) to analyse special classes of elliptic solutions of the Manakov system and its variants. The special case $R_j=\text{const.}$ is called a standing wave. We note that the complete classification of group-invariant solutions of the Manakov system \eqref{vNLS} in the case of two- and 3-coupled waves in $2+1$ dimensions  was performed in \cite{SciarrinoWinternitz1997}.

Substituting the ansatz \eqref{uni-wave} into \eqref{sys} and equating real and imaginary parts of the
resulting equations, we obtain
\begin{eqnarray}\label{red-vNLS}
  R_j''+p_j(x)R_j-R_j \theta_j'&=&\left(g_{j1}(x)R_1^2+g_{j2}(x)R_2^2\right)R_j, \quad \quad j=1,2 \\
   2 R _ { j } ^ { \prime } \theta _ { j } ^ { \prime } + R _ { j } \theta _ {j} ^ { \prime \prime } &=& 0,  \quad j=1,2,
\end{eqnarray}
where $p_j(x)=-\mu_j-V_j(x)$. Integrating once the second set of equations we find the first integrals (conservation of angular momentum)
\begin{equation}\label{phase}
\theta _ { j } ^ { \prime } ( x ) = \frac { c _ { j } } { R _ { j } ^ { 2 } ( x ) } , \quad j = 1,2,
\end{equation}
where $c_j$ are arbitrary constants. If $c_1=c_2=0$, then $\theta_1=\theta_2=\text{const}$, which can be assumed to be zero (trivial phase) by the gauge invariance of $\psi_i$. We note that the presence of nontrivial phases $c_j\ne 0$ implies nonzero current of the matter, which is proportional to $R^2_j(x) \theta_j^{ \prime }(x) dx=|\psi_j|^2\theta_j^{ \prime }(x) dx=c_j$ for each component $j=1,2$.

Using these first integrals, the  amplitude equations can be written as
\begin{equation}\label{ode-sys}
  R_j''+p_j(x)R_j=c_j^2 R_j^{-3}+(g_{j1}(x)R_1^2+g_{j2}(x)R_2^2)R_j,  \quad j=1,2.
\end{equation}

In the particular case of constant coefficients and trivial phase $c_j=0$, $j=1,2$, numerous exact solutions in terms of Jacobi elliptic functions have sporadically appeared in the literature \cite{PorubovParker1999, Hioe1997, FlorjanczykTremblay1989, Hioe1998,  Hioe2002, HioeCarroll2002, HioeSalter2002, ChristiansenEilbeckEnolskiiKostov2000, Kostov2007, KostovEnolskiiGerdjikovKonotopSalerno2004a}. In \cite{Hioe1999}, a hierarchy of exact analytic solitary-wave (doubly-periodic) solutions were presented in the case when the nonlinear coupling parameters $g_{ij}$ can change continuously and cover many regions. The ingenious idea of this  paper is to make the ansatz
\begin{equation}\label{ansatz}
  R_1(x)=\sqrt{C_1}f_p^{(n)}(\alpha x),  \quad R_2(x)=\sqrt{C_2}f_q^{(n)}(\alpha x),
\end{equation}
where $n=1,2$, $p,q=1,2,\ldots, 2n+1$, $p\leq q$ for $n=1$, and $p<q$ for $n=2$. $C_1$ and $C_2$ are required to be real and positive in order for the solutions to be real. Moreover, $f_j^{(n)}(u)$ are the Lam\'{e} functions, i.e., the eigenfunctions of the Lam\'{e} equation of order $n$
\begin{equation}\label{Lame}
  \frac{d^2f}{du^2}+[h-n(n+1)k^2\sn^2(u,k)]f=0,
\end{equation}
corresponding to the eigenvalues $h_j^{(n)}$, $j=1,2,\ldots, 2n+1$, arranged in descending
order of magnitude. Substitutions of the ansatz \eqref{ansatz} into Eqs. \eqref{ode-sys} for constant coefficients result in algebraic equations for the $h$'s, $E$'s, $C$'s, and $\alpha$ and $k^2$. The starting point is to express the square of the $j$th Lam\'{e} function of order $n$ in a power series in $s=\sn(u,k)$
$$\left[ f _ { j } ^ { ( n ) } ( u ) \right] ^ { 2 } = \sum _ { i = 1 } ^ { n + 1 } a _ { i j } ^ { ( n ) } s ^ { 2 ( i - 1 ) } , \quad j = 1 , \ldots , 2 n + 1.$$
There arose three types of waves of order one and of the same form and of order two, respectively, and ten types of order two for $R_j$.
All known exact analytic solutions that exist in the literature can be extracted from the results of \cite{Hioe1999}.

More recently, a new approach to construct exact solutions for the inhomogeneous case has been developed in \cite{Belmonte-BeitiaPerez-GarciaBrazhnyi2011}. The authors of this paper attempted to reduce the system to a single ODE whose solutions are apparent in terms of periodic and non-periodic functions.  In the present paper, building on the same idea, we will concentrate on the more general case of solutions with non-trivial phase $c_j\ne 0$.

\section{Lie point symmetry approach}

We apply the standard Lie symmetry algorithm to solve the amplitude equations \eqref{ode-sys}.
A vector field $V$ of the form
\begin{equation}\label{generic-vf}
  V=a(x,R_1,R_2)\gen x+b(x,R_1,R_2)\gen {R_1}+d(x,R_1,R_2)\gen {R_2}
\end{equation}
on the base jet space $\mathsf{J}^0(x,R_1,R_2)$ will be an element of the Lie point symmetry algebra $\lie$ of system \eqref{ode-sys} if
its second prolongation to the jet space $\mathsf{J}^2$ annihilates  the system on its solutions. This condition generates a set of equations to determine the coefficients $a,b,d$ of $V$. Equations independent of the coefficients of  \eqref{ode-sys} are solved to give
$$a_{xR_1}=a_{xR_2}=a_{R_1R_1}=a_{R_1R_2}=a_{R_2R_2}=0,$$
$$b_{R_1R_1}=b_{R_1R_2}=b_{R_2R_2}=0,$$
$$d_{R_1R_1}=d_{R_1R_2}=d_{R_2R_2}=0,$$
whose integration gives
$$a(x,R_1,R_2)=a_1 R_1+a_2 R_2+a(x),$$
$$b(x,R_1,R_2)=b_1(x) R_1+b_2(x) R_2+b(x),$$
$$d(x,R_1,R_2)=d_1(x) R_1+d_2(x) R_2+d(x),$$ where $a_1,a_2$ are arbitrary constants. The rest of determining equations provides $a_1=a_2=0$, $b=b_2=d=d_1=0$,
\begin{equation}\label{deteqs}
     b_1''+2p_1 a'+p_1'a=0, \quad  d_2''+2p_2 a'+p_2'a=0, \quad a''=2b_1'=2d_2',
\end{equation}
and equations involving $g_{ij}$.
From the last two equations we obtain
$$b_1(x)=\frac{1}{2}a'(x)+k_1, \quad   d_2(x)=\frac{1}{2}a'(x)+k_2,$$
where $k_1,k_2$ are integration constants. The requirement $c_j\ne 0$ imposes $k_j=0$, $j=1,2$.
Finally we find that  the Lie symmetry algebra of \eqref{ode-sys} is represented by vector fields of the form
\begin{equation}\label{vf-a}
  V=a(x)\gen x+\frac{a'(x)}{2}(R_1\gen{R_1}+R_2\gen{R_2}),
\end{equation}
where $a(x)$ satisfies the set of  linear differential equations
\begin{equation}\label{3rd-eq}
  \mathbb{M}_j(a)=a'''+4p_ja'+2p_j'a=0, \quad j=1,2
\end{equation}
and the arbitrary coefficients $g_{ij}(x)$ are related to $a(x)$ by the relation
\begin{equation}\label{coeff}
  g_{ij}(x)=h_{ij}a(x)^{-3}.
\end{equation}
The set of ODEs \eqref{3rd-eq} admits the first integrals
\begin{equation}\label{two-first-int}
  E_j=\frac{1}{4}(2aa''-a'^2)+p_j(x)a^2,  \quad j=1,2,
\end{equation}
from which it follows that the potentials are related by
\begin{equation}\label{p-a}
  K\equiv E_1-E_2=(p_1-p_2)a^2.
\end{equation}
When $p_1=p_2=p$ ($E_1=E_2=E$), $a(x)\ne 0$ satisfies a single third-order equation
\begin{equation}\label{3rd-eq-bis}
a'''+4pa'+2p'a=0.
\end{equation}
Integrating $V$, we get the one-parameter symmetry group transformation as
\begin{equation}\label{sym-group}
  \tilde{x}=F^{-1}(F(x)+\varepsilon),  \quad \tilde{R}_j=\sqrt{\frac{a(\tilde{x})}{a(x)}}R_j,  \quad F(x)=\int^x \frac{ds}{a(s)}.
\end{equation}
From the relationships
$$F(\tilde{x})=F(x)+\varepsilon, \qquad \frac{\tilde{R}_j}{\sqrt{a(\tilde{x})}}=\frac{R_j}{\sqrt{a(x)}},$$ it immediately follows that $U_j=a(x)^{-1/2}R_j$ (invariants) and $y=F(x)$ are the canonical (or normal) coordinates   of the symmetry group and in these coordinates the symmetry group is simply a translational one along $y$.

For given $p$, the general solution to this equation is $a(x)=A u^2+B uv+C v^2$,
where $u$, $v$ are the linearly independent solutions of the Hill's equation $f''+pf=0$.
The point transformation
\begin{equation}\label{recti}
  y=\int^x \frac{ds}{a(s)},  \quad U_j(y)=\frac{R_j}{\sqrt{a}}
\end{equation}
rectifies the vector field   $V$ to $\widehat{V}=\gen y$. Taking into account the first integrals \eqref{two-first-int}, this transformation maps the system \eqref{ode-sys} to the autonomous one
\begin{equation}\label{autonomous-sys}
  U_j''+E_j U_j=c_j U_j^{-3}+(h_{j1}U_1^2+h_{j2}U_2^2)U_j,  \quad j=1,2.
\end{equation}

Now, our objective is to identify exact solutions of the autonomous system \eqref{autonomous-sys} and then to revert the change \eqref{recti}, thus obtaining exact solutions of the original system \eqref{ode-sys} with inhomogeneous coefficients.

\section{Reduction to an integrable singular ODE}
Under the condition $E_1=E_2=E$, we assume a set of  solutions which are proportional among themselves
\begin{equation}\label{sol-set}
  U_i(y)=\delta_i \Phi(y),  \quad i=1,2
\end{equation}
such that $\Phi$ satisfies the following ODE for some $c\ne 0$ and $\sigma\ne 0$
\begin{equation}\label{single-ode}
  \Phi''+E \Phi=c \Phi^{-3}+2\sigma \Phi^3.
\end{equation}
Introducing \eqref{sol-set} into the system \eqref{autonomous-sys} and using Eq. \eqref{single-ode} we find that we must have $\delta_1^4={c_1}/{c}$, $\delta_2^4={c_2}/{c}$ and the following relations should be satisfied among the parameters $\delta_i$, $c$ and $\sigma$
\begin{equation}\label{algebraic-sys}
  \begin{split}
     & c_1 h_{11}+\sqrt{c_1c_2}h_{12}=2\sigma \sqrt{cc_1},  \\
      & \sqrt{c_1c_2}h_{21}+c_2 h_{22}=2\sigma \sqrt{cc_2}.
  \end{split}
\end{equation}
For the consistency of constraints \eqref{algebraic-sys} we need to impose the condition
\begin{equation}\label{compatible}
  c_1(h_{11}-h_{21})^2-c_2(h_{22}-h_{12})^2=0.
\end{equation}
In the special case $h_{11}=h_{21}$, $h_{22}=h_{12}$,  this condition is satisfied for any $c_1,c_2$.

Equation \eqref{single-ode} with $c=0$ is the familiar Duffing equation with cubic nonlinearity and was profusely commented in \cite{Belmonte-BeitiaPerez-GarciaBrazhnyi2011}. When, $c\ne 0$, the equation presents a spatial singularity in the origin and the phase plane changes completely. We comment that this equation passes the Painlev\'{e} test (a necessary condition for it to have the Painlev\'{e} property) for any values of $E,c, \sigma$. It does indeed have the Painlev\'{e} property. This implies that, as we will see, the  solution of \eqref{single-ode} can  be obtained in terms of Jacobi elliptic functions. Indeed, \eqref{single-ode} can be integrated once and provides the first integral
\begin{equation}\label{first-int-3}
  \Phi'^2=\sigma\Phi^4-c\Phi^{-2}-E\Phi^2+C_0.
\end{equation}
This equation belongs to the general class $\Phi'^2=F(y,\Phi)$ with $F$ being rational in $\Phi$ and analytic in $y$. If such an equation possesses  the Painlev\'{e} property it should be equivalent either to the Riccati equation or to the equation for the Jacobi elliptic functions. This is the case here. The transformation $\Phi=Q(y)^{-1/2}$, $Q(y)>0$ suggested by the Painlev\'{e} analysis transforms \eqref{first-int-3} to the elliptic function equation
\begin{equation}\label{elliptic-eq0}
  Q'^2=P(Q)=4Q(-c Q^3+C_0 Q^2-E Q+\sigma).
\end{equation}

Instead of dealing with \eqref{elliptic-eq0}
we  rather use another transformation $Q=W^{-1}$ (a type of M\"obius transformation)  taking \eqref{first-int-1} to a special elliptic function equation
\begin{equation}\label{elliptic-eq}
  W'^2=4(\sigma W^3-E W^2+C_0 W-c)=4\sigma(W-W_1)(W-W_2)(W-W_3).
\end{equation}
Eqs. \eqref{first-int-3} and \eqref{elliptic-eq} are linked by the transformation $\Phi=W^{1/2}$.
Note that $\sigma$ can be normalized to $\pm 1$ by a scaling of the independent variable $y\to \sqrt{|\sigma|}y$. First, we consider the case $\sigma>0$.  With the change  $y\to \sqrt{|\sigma|}y$, the resulting equation is
\begin{equation}\label{elliptic-eq2}
  W'^2=4(W^3-\frac{E}{\sigma} W^2+\frac{C_0}{\sigma} W-\frac{c}{\sigma})=4(W-W_1)(W-W_2)(W-W_3),
\end{equation}
where $W_1,W_2,W_3$ are the roots of the cubic polynomial (that can be complex or real) on the right side, which are uniquely determined by the relations
\begin{equation}\label{constraints}
W_1+W_2+W_3=\frac{E}{\sigma},\quad W_1W_2+W_2W_3+W_1W_3=\frac{C_0}{\sigma},\quad W_1W_2W_3=\frac{c}{\sigma}.
\end{equation}
$W$ needs to be real and nonnegative because $\Phi=\sqrt{W}$. For $W_1\ne W_2\ne W_3$, the general solution of \eqref{elliptic-eq2} can be expressed in terms of doubly-periodic Jacobi elliptic functions. This is achieved by a transformation of the form $Z^2=M(W)=(a W+b)/(c W+d)$, $ad-bc>0$, for  suitably chosen M\"obius maps in $W$ (these maps are known to preserve cross ratios) and also for special choice of $\lambda$, $k$ (modulus) in terms of $W_i$, into the standard form of the equation for the Jacobi elliptic function $Z=\sn(\lambda y,k)$
\begin{equation}\label{standard-Jacobi}
  Z'^2=(1-Z^2)(1-k^2 Z^2),  \quad 0\leq k\leq 1
\end{equation}
such that $W=W_1$, $W=W_2$, $W=W_3$ are mapped to $Z^2=1$, $Z^2=k^{-2}$, $Z^2=\infty$, respectively. The case of multiple roots gives rise to  solutions in terms of elementary functions.

In light of the previous observations, we find that real solutions can be finite if all three roots $W_i$ are positive.

If $0<W_1\leq W\leq W_2<W_3$, then
\begin{equation}\label{finite-sol-W}
 W=W_1+(W_2-W_1)\sn^2(\lambda y,k),  \quad \lambda^2=W_3-W_1,  \quad k^2=\frac{W_2-W_1}{W_3-W_1},
\end{equation}
where $\sn(u,k)$ is a Jacobi elliptic function with the property $\sn(u,0)=\sin u$, $\sn(u,1)=\tanh u$. The appropriate M\"obius map is given by $M(W)=(W-W_1)/(W_2-W_1)$ with $\lambda$ and $k$ as in \eqref{finite-sol-W}.

If any two of the roots $W_{1,2,3}$ coincides, the solutions degenerate to elementary functions of trigonometric or hyperbolic type. For example, in the limit $W_3\to W_2$ ($k=1$) this solution reduces to a soliton (in Optics, sometimes called a "dark soliton")
\begin{equation}\label{soliton}
  W=W_1+(W_2-W_1)\tanh^2(\sqrt{W_2-W_1}(y-y_0))=W_2-\frac{W_2-W_1}{\cosh^2(\sqrt{W_2-W_1}(y-y_0))}.
\end{equation}

There are also real nonnegative singular solutions of \eqref{elliptic-eq2}, exhibiting a blow-up. When the roots are all distinct and ordered as $0<W_1<W_2<W_3\leq W$, then we have the singular solution
\begin{equation}\label{singular-sol-W}
  W=W_1+\frac{W_3-W_1}{\sn^2(\lambda y,k)}, \quad \lambda^2=W_3-W_1,  \quad k^2=\frac{W_2-W_1}{W_3-W_1}.
\end{equation}
For $W_2=W_1$, this solution reduces to an elementary singular periodic one, and to a singular soliton if $W_3=W_2$.
It is difficult to attribute a direct physical meaning to these types of singular periodic or nonperiodic solutions.

The case of $\sigma<0$ can be discussed similarly. In this case real finite solutions are obtained
if $W_1<0<W_2\leq W\leq W_3$ in the form
\begin{equation}\label{sn_sigma_minus}
W=W_3-(W_3-W_2)\sn^2(\lambda y,k),  \quad \lambda=\sqrt{W_3-W_1},  \quad k^2=\frac{W_3-W_2}{W_3-W_1}.
\end{equation}

In the limit $W_2\to W_1$ we have
\begin{equation}\label{soliton2}
W=W_1+\frac{W_3-W_1}{\cosh^2(\sqrt{W_3-W_1}(y-y_0))}.
\end{equation}

The solutions of \eqref{elliptic-eq2} can be related to the Weierstrass function $\wp(y)\equiv \wp(y;g_2,g_3)$ with invariants $g_2, g_3$. First we note that the cubic polynomial on the right hand side of \eqref{elliptic-eq2} can be changed to one with no quadratic power (called reduced cubic) by a simple shift of $W$: $W \to W+E/(3\sigma)$.  The defining equation  for $\wp(y)$ is
\begin{equation}\label{Wei}
  \left\{\wp  ' ( y ) \right\} ^ { 2 } = 4 \wp ^ { 3 } - g _ { 2  }\wp - g _ { 3 }=4(\wp-e_1)(\wp-e_2)(\wp-e_3),
\end{equation}
with the following relations between the invariants $g_2$ and $g_3$ and the roots $e_1, e_2, e_3$ of the cubic polynomial $4y^3-g_2y-g_3$
\begin{equation}\label{g23-roots}
  e_1+e_2+e_3=0,  \quad  e_1e_2+e_1e_3+e_2e_3=-\frac{g_2}{4},  \quad e_1e_2e_3=\frac{g_3}{4}.
\end{equation}
From the first two relations it follows that $g_2$ must be nonnegative, $g_2\geq 0$. If we assume $W_i=e_i$, $i=1,2,3$, then $W(y)$ is identified with $\wp(y)$.

If $g_2g_3\ne 0$, there is a relationship between $k^2=(e_2-e_1)/(e_3-e_1)$ and the invariants $g_2$, $g_3$ \cite{Davis1962}
\begin{equation}\label{k-invar-rel}
  108g_3^2(1-k^2+k^4)^3=g_2^3(k^2+1)^2(k^2-2)^2(2k^2-1)^2.
\end{equation}
The discriminant of the cubic equation $4y^3-g_2y-g_3=0$ is defined by
\begin{equation}\label{discr}
  \Delta=16\prod_{i<j}(W_i-W_j)^2=g_2^3-27 g_3^2.
\end{equation}
The roots are real if and only if $\Delta \geq 0$. There is a complex root if $\Delta<0$. At least two roots are equal if $\Delta=0$.
If $g_2=0$ and $g_3\ne 0$, then $\Delta<0$; one root is real and the other two are complex conjugates and $k^2$ satisfies $k^4-k^2+1=0$. If, on the other hand, $g_2\ne 0$ ($>0$) and $g_3=0$, then $\Delta>0$; the roots are all real and $k^2\in\curl{-1,1/2,2}$.

Once a particular form of $W$ is selected, the amplitude coordinates of the original inhomogeneous system are easily obtained from \eqref{sol-set} and \eqref{recti}, taking the form
$$
R_j(x)=\delta_j \sqrt{a(x)W\left(\int^x\frac{ds}{a(s)}\right)},
$$
while the phase component is obtained by a simple integration in \eqref{phase}.

\section{Exact solutions for the case $h_{ij}=1$}

In this section, for the completeness of the analysis,  we will study system \eqref{autonomous-sys} in the particular case that $E_1=E_2=E$  and $h_{ij}=1$ ($a(x)=1$) for $i,j=1,2$. In this case,  switching to polar coordinates $(U,\gamma)$ defined by
\begin{equation}\label{polar}
  U_1=U \cos \gamma,  \quad U_2=U \sin \gamma.
\end{equation}
allows us to decouple the system
\begin{equation}\label{hij=1}
  U_j''+E U_j=c_j U_j^{-3}+(U_1^2+U_2^2)U_j, \quad j=1,2.
\end{equation}
In Ref. \cite{SciarrinoWinternitz1997}, the idea of introducing polar or spherical coordinate representations for some particular ODE systems obtained by symmetry reduction for the system \eqref{vNLS} in the case of two- or three-component waves also made the decoupling possible.

From system \eqref{hij=1} we can write
\begin{equation}\label{comb1}
  U_2 U_1''-U_1 U_2''=\frac{d}{dy}[U_1' U_2-U_1 U_2']=-(\gamma' U^2)'=c_1 U_1^{-3}U_2-c_2U_2^{-3}U_1.
\end{equation}

This equation can be integrated once by using the integrating factor $2U^2\gamma'$
\begin{equation}\label{first-int-1}
  U^4\gamma'^2+\frac{c_1^2}{\cos^2 \gamma}+\frac{c_2^2}{\sin^2 \gamma}=K_1.
\end{equation}
Again from \eqref{hij=1} we find
\begin{equation}\label{comb2}
  U_1 U_1''+U_2 U_2''+E U^2=c_1 U_1^{-2}+c_2 U_2^{-2}+U^4.
\end{equation}
Taking into account the relation
$$U_1 U_1''+U_2 U_2''=UU''-\gamma'^2 U^2$$ and writing the first integral \eqref{first-int-1} in the form
$$U^2 \gamma'^2=K_1 U^{-2}-c_1 U_1^{-2}-c_2 U_2^{-2}$$
we can decouple the equation for $U$
\begin{equation}\label{U-eq}
  U''+EU=U^3+K_1 U^{-3}.
\end{equation}
This equation has the same form as Eq. \eqref{single-ode}. Thus, we can integrate it as before and its first integral is
\begin{equation}\label{first-int-2}
  U'^2=\frac{1}{2}U^4-E U^2-K_1U^{-2}+K_2.
\end{equation}
By the scaling $y\to y/\sqrt{2}$, we can replace it with
\begin{equation}\label{first-int-2-0}
  U'^2=U^4-2E U^2-2K_1U^{-2}+2K_2.
\end{equation}
The substitution
\begin{equation}\label{subst}
  U(y)=\sqrt{W(y)}
\end{equation}
transforms \eqref{first-int-2} to an elliptic function equation
$$
  W'^2=4(W-W_1)(W-W_2)(W-W_3),
$$
which is exactly eq. \eqref{elliptic-eq2}, in this case with $\sigma=1$ and
$$W_1+W_2+W_3=2E,  \quad W_1W_2+W_2W_3+W_3W_1=2K_2,  \quad W_1W_2W_3=2K_1.$$
Therefore, all the discussion done in Section 3 is valid.

Now we can turn to solve the phase equation \eqref{first-int-1} for $\gamma$. We change $\gamma$ to $\gamma(y)=\Omega(\zeta(y))$ with the introduction of a new coordinate
$$\zeta(y)=\int^y \frac{dx}{U^2(x)}$$ and rewrite \eqref{first-int-1} as
$$\Omega'^2+\frac{c_1^2}{\cos^2 \gamma}+\frac{c_2^2}{\sin^2 \gamma}=K_1>0.$$
A further change of the dependent variable $\phi(\zeta)=\sin \Omega$ simplifies it to the separable ODE
$$\phi'^2+\frac{c_1^2 \phi^2+(1-\phi^2)(c_2^2-K_1 \phi^2)}{\phi^2}=0.$$
Separating variables we find
$$\frac{\phi d\phi}{\sqrt{(\phi^2-1)(c_2^2-K_1 \phi^2)-c_1^2\phi^2}}=d\zeta,$$
or
$$\frac{d\tau}{\sqrt{-K_1\tau^2+(K_1-c_1^2+c_2^2)\tau-c_2^2}}=2d\zeta$$
after the substitution $\tau=\phi^2$. Carrying out the integration gives us
$$\phi^2=\frac{1}{2K_1}\left[b+\sqrt{\Delta}\sin 2\sqrt{K_1}(\zeta-\zeta_0)\right],$$
and
\begin{equation}\label{phase_polar}
\gamma(y)=\Omega(\zeta)=\arcsin\left[\frac{1}{\sqrt{2K_1}}\left(b+\sqrt{\Delta}\sin 2\sqrt{K_1}(\zeta-\zeta_0)\right)^{1/2}\right],
\end{equation}
where
$$K_1>0,  \quad b=K_1-c_1^2+c_2^2, \quad \Delta=b^2-4K_1 c_2^2.$$
This completes the integration of the system.

Thus, in summary, Eqs. \eqref{polar} correspond to the solution of system \eqref{autonomous-sys} with $U$ and $\gamma$  given by Eqs. \eqref{subst} and \eqref{phase_polar}, respectively, and where $W$ corresponds to the solutions \eqref{finite-sol-W}, \eqref{soliton} or \eqref{singular-sol-W}. Then, we obtain analytical solutions for the nonlinear Schr\"odinger system with non-trivial phase.

It is remarkable to note that the nonlinearities $g_{ij}$ are constants for this case and therefore, the method also allows us construct solutions for ``standard'' nonlinear Schr\"odinger coupled systems.

\section{Physical Applications}

In this section, we consider some examples which are relevant in  physical applications corresponding to Feschbach resonance management for Bose--Einstein condensates.

\subsection{Systems without external potential ($V_1(x)=V_2(x)=0$) and non-trivial phase ($\theta(x)\neq 0$)}

As a first application of our ideas, we consider first a system without external potentials $V_1(x)=V_2(x)=0$. Taking $\mu_1=\mu_2\equiv\mu$, then $p(x)$ is constant with value $-\mu$, and the general solution of
\eqref{3rd-eq-bis} is
\begin{eqnarray}
    a(x)&=&C_1 \sin \omega x+C_2\cos \omega x+C_3,\ (\mu<0) , \label{eq1}\\
     a(x)&=&C_1 e^{\omega x}+C_2 e^{ -\omega x}+C_3,\ (\mu>0),\label{eq2}
\end{eqnarray}
where $\omega=2\sqrt{|\mu|}$. Note that the Hill's equation associated to \eqref{3rd-eq-bis} is $f''-\mu f=0$. Now the coefficients are given by \eqref{coeff}. The transformation \eqref{recti} passes the system to the autonomous form \eqref{autonomous-sys}, where we have the exact solutions identified in Section 3 and 4.

Let us consider the case $\mu=-\omega^2/4<0$. Then, taking $C_1= 0$, $C_2 =\alpha$ and $C_3= 1$, Eq. \eqref{eq1} leads to a periodic dependence of the coefficient, $a(x)=1+\alpha \cos\omega x$ ($E=\omega^2(1-\alpha^2)/4$) and according to Eq. \eqref{coeff} the nonlinear coefficients are given by
$$
g_{ij}=\frac{h_{ij}}{(1+\alpha \cos\omega x)^3}, \quad i,j=1,2.
$$
This nonlinear coefficient has a direct physical interest, since in the limit for small $\alpha$, this nonlinearity is approximately harmonic in space, $g_{ij}=h_{ij}(1-3\alpha\cos\omega x)$, $\alpha\ll 1$, $i,j=1,2$.

Using formula \eqref{recti}, we obtain the following equation for $y(x)$:
\begin{equation}\label{transf_y}
\tan\left(\frac{\omega}{2}\sqrt{1-\alpha^2}y(x)\right)=\sqrt{\frac{1-\alpha}{1+\alpha}}\tan\frac{\omega x}{2}.
\end{equation}
Thus, using equation \eqref{soliton}, we can construct dark-dark soliton solutions of Eqs. \eqref{sys} for $V_{1}=V_{2}=0$, $g_{ij}(x)=h_{ij}/(1+\alpha \cos\omega x)^3$, $i,j=1,2$ with
$$
\psi_j(t,x)=R_j(x)e^{i(\theta_j(x)+\mu t)},\quad j=1,2
$$
and where
\begin{equation}\label{sol_dark_dark}
R_j(x)=(1+\alpha \cos\omega x)^{1/2}U_j(y(x)),\quad j=1,2
\end{equation}
with  $U_j$ being as
$$
U_j(y(x))=\delta_j \sqrt{W_1+(W_2-W_1)\tanh^2(\sqrt{W_2-W_1}y(x))},\quad j=1,2,
$$
where $\delta_j=(c_j/c)^{1/4}$, $j=1,2$, being $c_j$ and $c$ given by Eqs. \eqref{algebraic-sys} and $y(x)$ given by \eqref{transf_y}.

\begin{figure}
\begin{center}
\vspace{-4cm}
\includegraphics[height=13cm,width=6cm]{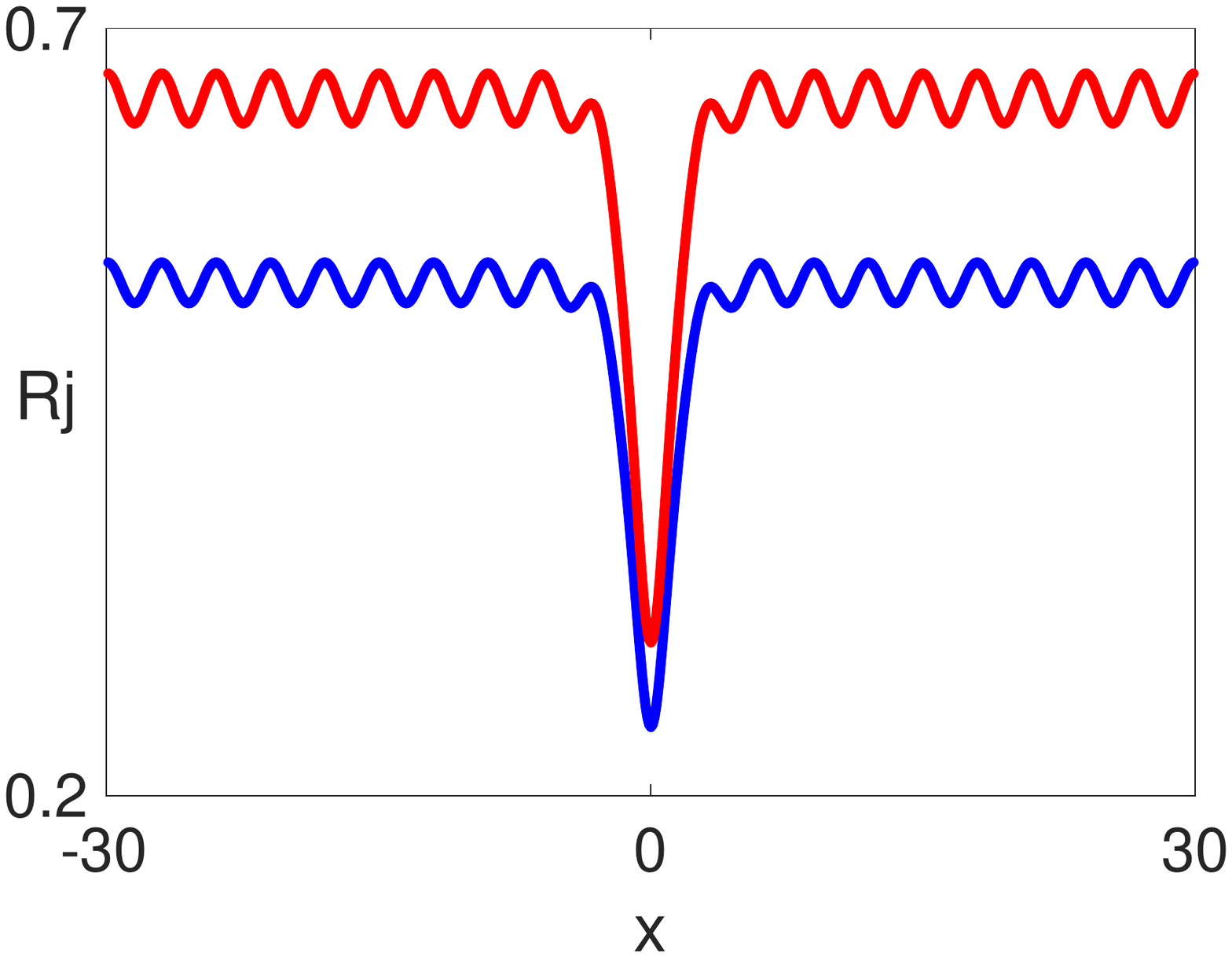}
\includegraphics[height=13cm,width=6cm]{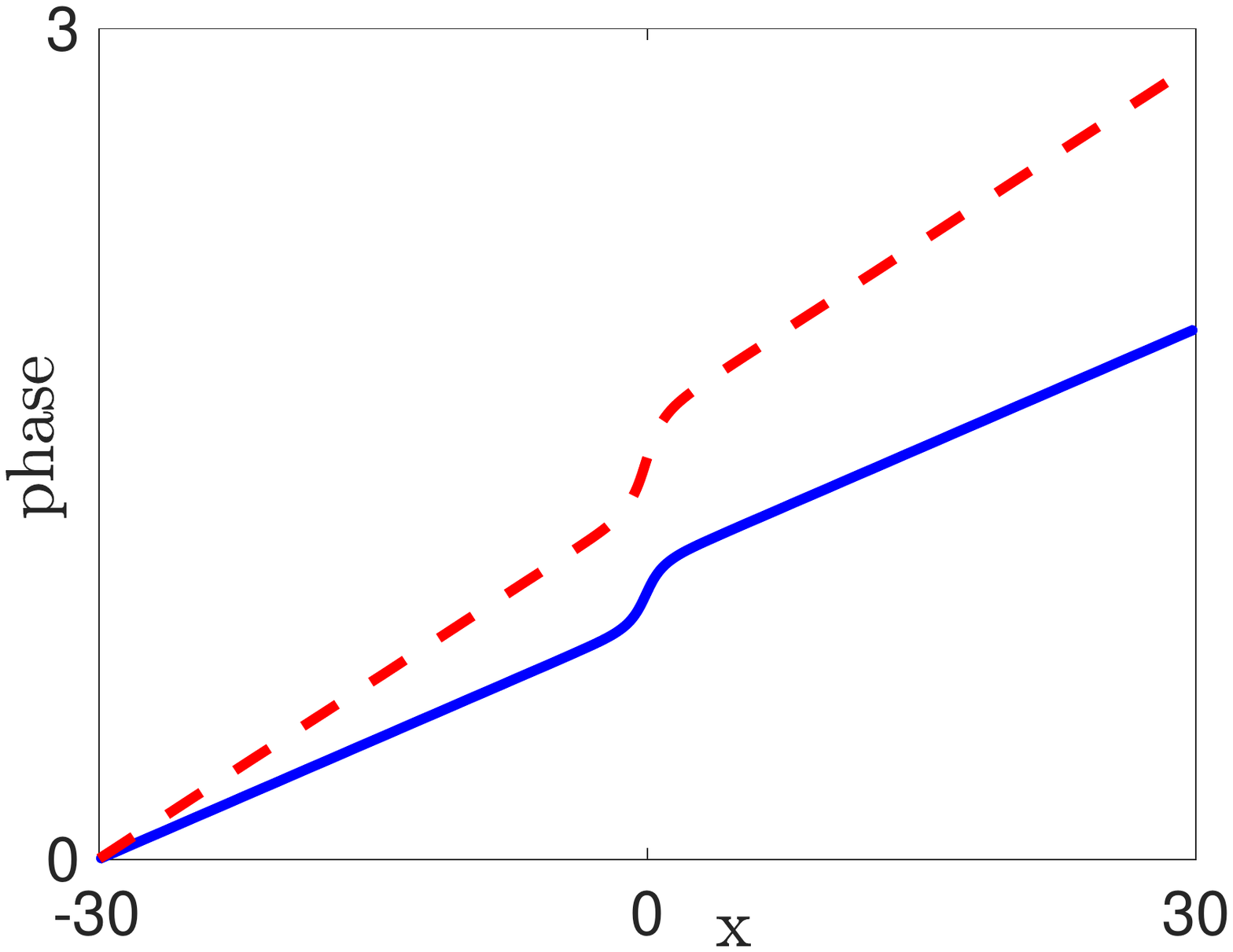}
\vspace{-3.5cm}
\caption{(Left) Dark-dark solution corresponding to \eqref{sol_dark_dark}. Blue solid curve corresponds to the first component and red solid curve corresponds to the second component (Right) Non-trivial phase $\theta_j$ (see Eq. \eqref{phase_dark_dark}). Blue solid curve corresponds to the first component and red dashed curve corresponds to the second component. The parameters used in the calculations can be found in the text.}\label{fig_dark_dark}
\end{center}
\end{figure}

Moreover, the phase $\theta_j(x)$ is given by
\begin{equation}\label{phase_dark_dark}
\theta_j(x)=\int^x \frac{c_j}{\delta_j^2(1+\alpha \cos\omega s)\left(W_1+(W_2-W_1)\tanh^2(\sqrt{W_2-W_1}y(s))\right)}ds.
\end{equation}
Fig. \ref{fig_dark_dark} shows the dark-dark solution \eqref{sol_dark_dark} together with the nontrivial phase \eqref{phase_dark_dark}.

These solutions were calculated using the parameters
\begin{equation}
\begin{split}
h_{11}&=2,\ h_{12}=1,\ h_{21}=1/2,\ h_{22}=2,\\
\alpha&=0.05,\ W_1=0.1,\ W_2=0.5,\ W_3=0.5,\ \sigma=1,\\
c&=0.025,\ m_s=\frac{h_{22}-h_{12}}{h_{11}-h_{21}}, \ c_2=\frac{4\sigma^2 c}{(m_s h_{11}+h_{12})^2},\ c_1=m_s^2c_2,\\
\delta_1&=\left(\frac{c_1}{c}\right)^{1/4},\ \delta_2=\left(\frac{c_2}{c}\right)^{1/4}.
\end{split}
\end{equation}

\begin{figure}
\begin{center}
\vspace{-4cm}
\includegraphics[height=12cm,width=10cm]{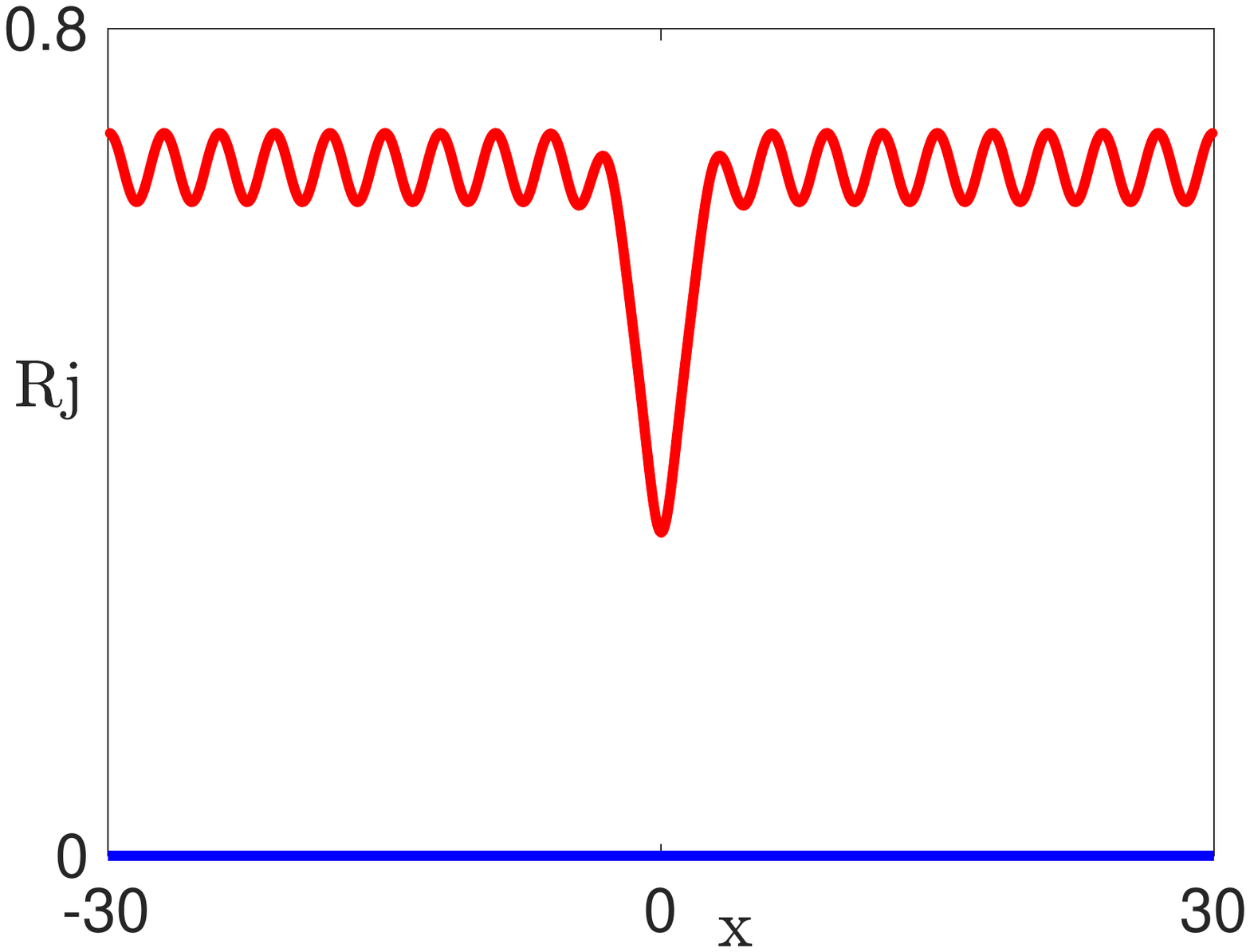}
\vspace{-3.5cm}
\caption{Zero-dark solution corresponding to \eqref{sol_dark_dark}. Blue solid curve (zero solution) corresponds to the first component and red solid curve (dark solution) corresponds to the second component. }\label{fig_dark_zero}
\end{center}
\end{figure}

On the other hand, it is interesting to note that it is possible to calculate a solution with a non-trivial phase in only one of its components, the other component being null. Thus, the method presented here allows to calculate a solution with a non-trivial phase both for a scalar equation or for a system of equations.  This fact is shown in Figure \ref{fig_dark_zero}. The parameters used were, for this case, $h_{11}=2,\ h_{12}=2,\ h_{21}=1/2,\ h_{22}=2,\ \alpha=0.1,\ W_1=0.1,\ W_2=W_3=0.5$.

\subsection{Systems with quadratic potentials $(V_1(x)=V_2(x)=\mu^2x^2)$  and non-trivial phase ($\theta(x)\neq 0$) }

Other possibility is to choose quadratic trapping potentials of the form $V_1(x)=V_2(x)=\mu^2x^2$, which are particularly relevant in applications. If the chemical potential is $\mu_1=\mu_2\equiv\mu$, ($p_1=p_2=-\mu^2 x^2-\mu$) we find that a solution of Eq. \eqref{3rd-eq-bis} is $a(x)=e^{\mu x^2}$ ($E=0$). Thus, the nonlinear terms for this case are
\begin{equation}
g_{ij}(x)=h_{ij}e^{-3\mu x^2},\quad i,j=1,2.
\end{equation}
Using formula \eqref{recti}, we obtain the following equation for $y(x)$:
\begin{equation}
y(x)=\frac{1}{2}\sqrt{\frac{\pi}{-\mu}}\text{erf}\left(\sqrt{-\mu}x\right)
\end{equation}
This equation leads us to a restriction on the sign of the chemical potential $\mu$ which must be negative, $\mu<0$.

Note that the range of $y(x)$ is finite where $-\frac{1}{2}\sqrt{\frac{\pi}{-\mu}}\leq y\leq \frac{1}{2}\sqrt{\frac{\pi}{-\mu}}$.

Since $E = 0$, using the first relation of \eqref{constraints}, at least one of the constants $W_i, i=1,2,3$ must be negative. Assume that $W_1<0$ and $W_2, W_3>0$.  Then, we are in the case where $\sigma <0$, and therefore, the solution for $W$ is given by Eq. \eqref{sn_sigma_minus} or alternatively, $W=\wp(y-y_0;g_2,g_3)$ with the invariants $g_2=-4C_0/\sigma\geq 0$, $g_3=4c/\sigma$. We find from \eqref{k-invar-rel} that given $W_1, W_3$ (or $k^2<1$), and $\sigma=-1$, then the parameters $C_0\geq 0$ and $c$ are constrained by the relation
\begin{equation}\label{rel-c-C0}
  27c^2(1-k^2+k^4)^3=C_0^3(k^2+1)^2(k^2-2)^2(2k^2-1)^2.
\end{equation}
For the special choice $k^2=1$, \eqref{rel-c-C0} gives $4C_0^3-27c^2=0$, which is equivalent to the vanishing of the discriminant \eqref{discr}, the solutions are reduced to hyperbolic functions.

Thus, the solutions for $U_j(y(x))$ are given by
\begin{equation}\label{sol_U}
U_j(y(x))=\delta_j \sqrt{W_3-(W_3-W_2)\sn^2(\lambda y(x),k)}, \quad j=1,2,
\end{equation}
with $\lambda^2=W_3-W_1$ and $k^2=(W_3-W_2)/(W_3-W_1)$.

\begin{figure}
\begin{center}
\vspace{-4cm}
\includegraphics[height=13cm,width=6cm]{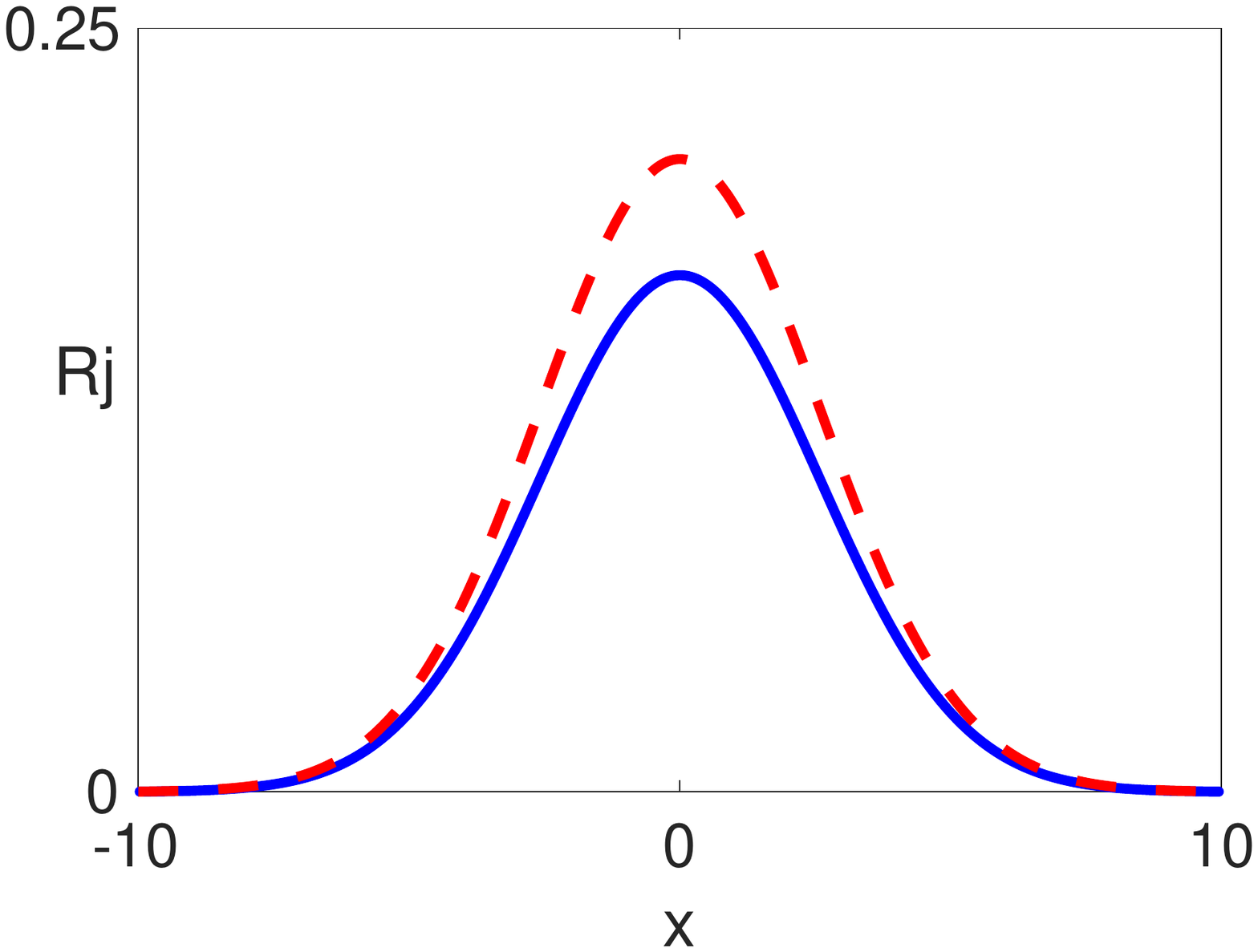}
\includegraphics[height=13cm,width=6cm]{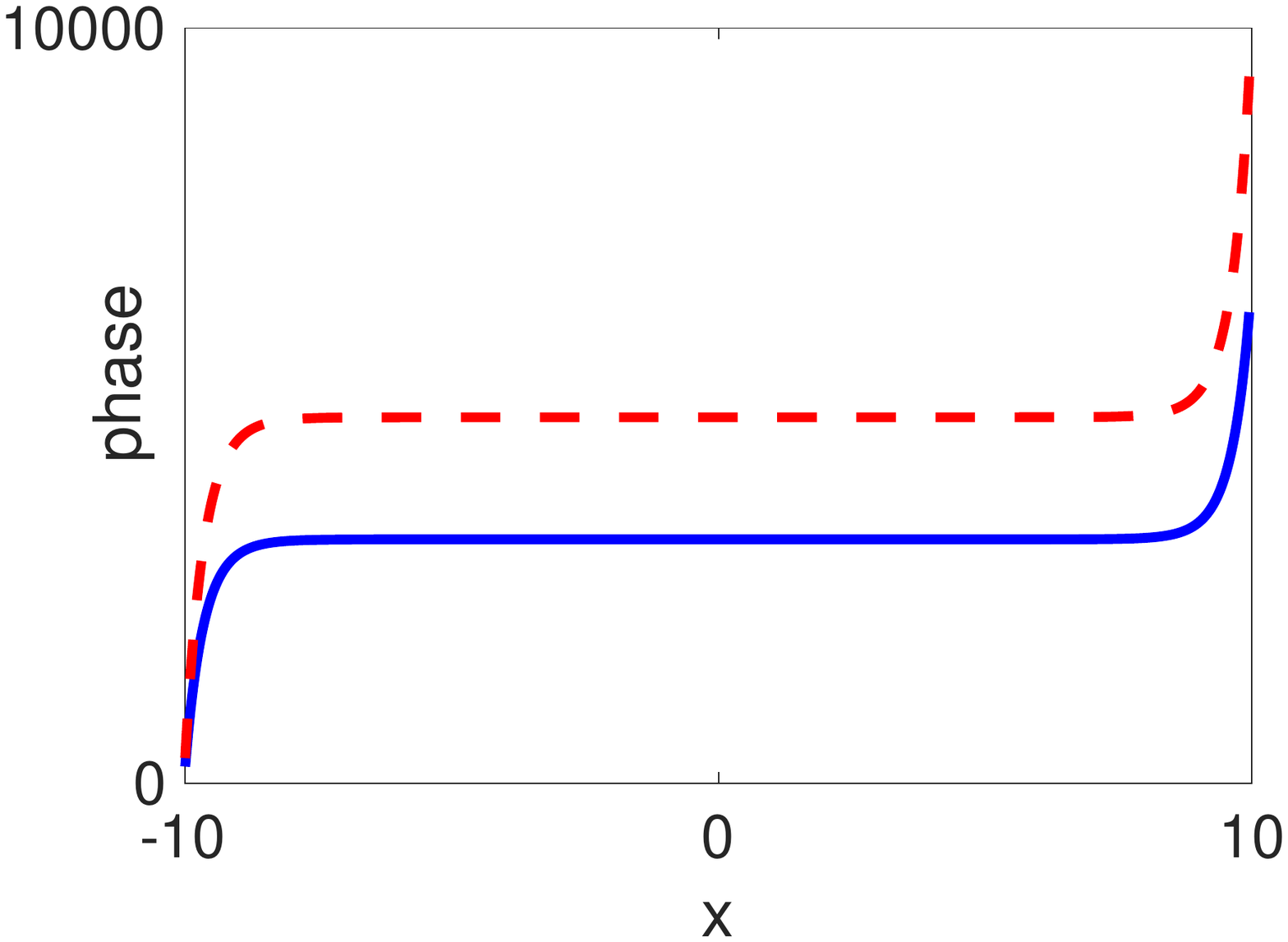}
\vspace{-3cm}
\caption{(Left) bright-- bright soliton solution calculated for the parameters $h_{11}=2, h_{12}=1, h_{21}=1/2, h_{22}=2, W_1=-0.1, W_3=0.0501, \mu=-0.15, \sigma=-1$ and $k^*=0.036$. Blue solid curve corresponds to the first component and red dashed curve corresponds to the second component (Right) non-trivial phase $\theta_j$ (see Eq. \eqref{phase_multi_peak}). Blue solid curve corresponds to the first component and red dashed curve corresponds to the second component.}\label{fig_bright_bright}
\end{center}
\end{figure}

\begin{figure}
\begin{center}
\vspace{-4cm}
\includegraphics[height=13cm,width=6cm]{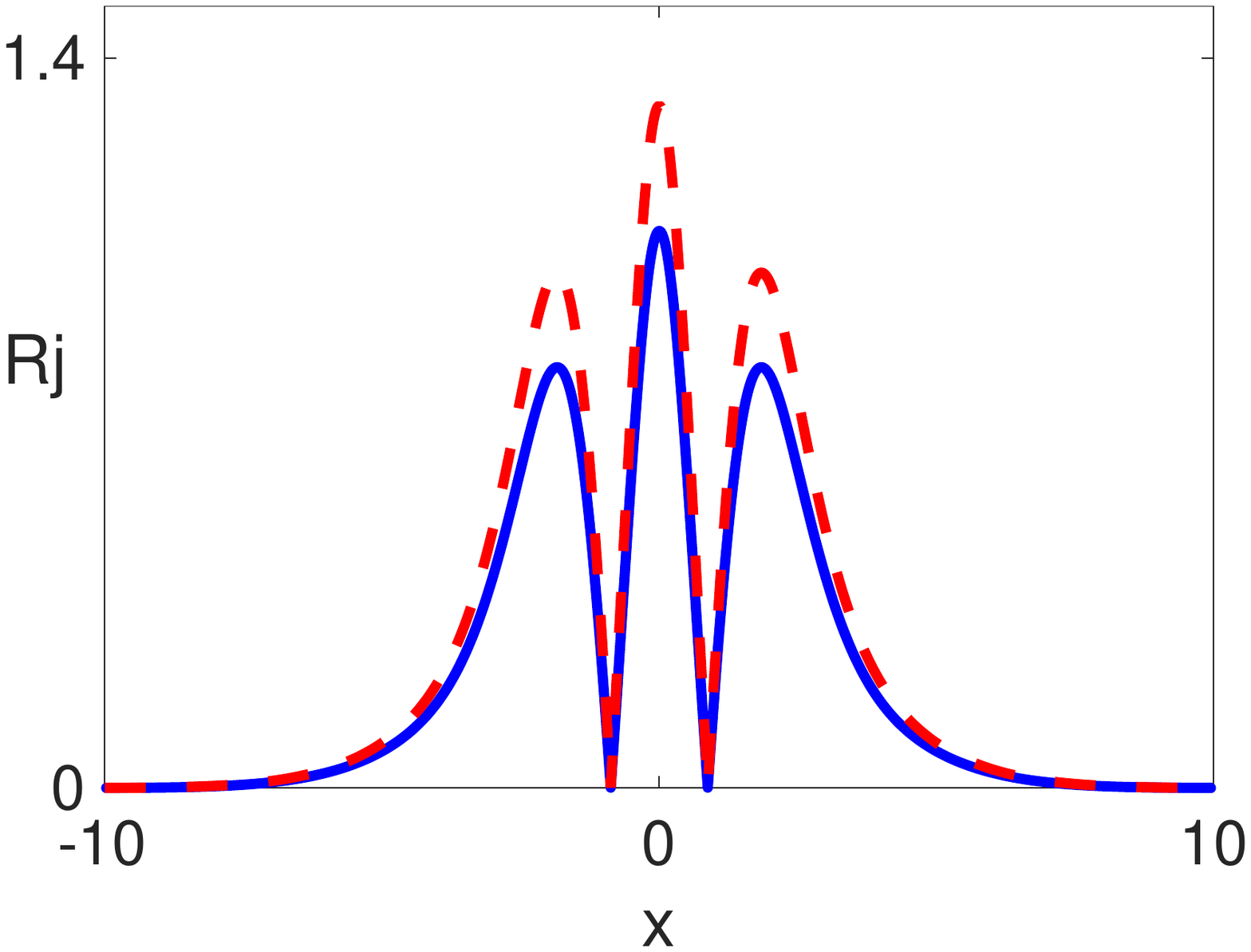}
\includegraphics[height=13cm,width=6cm]{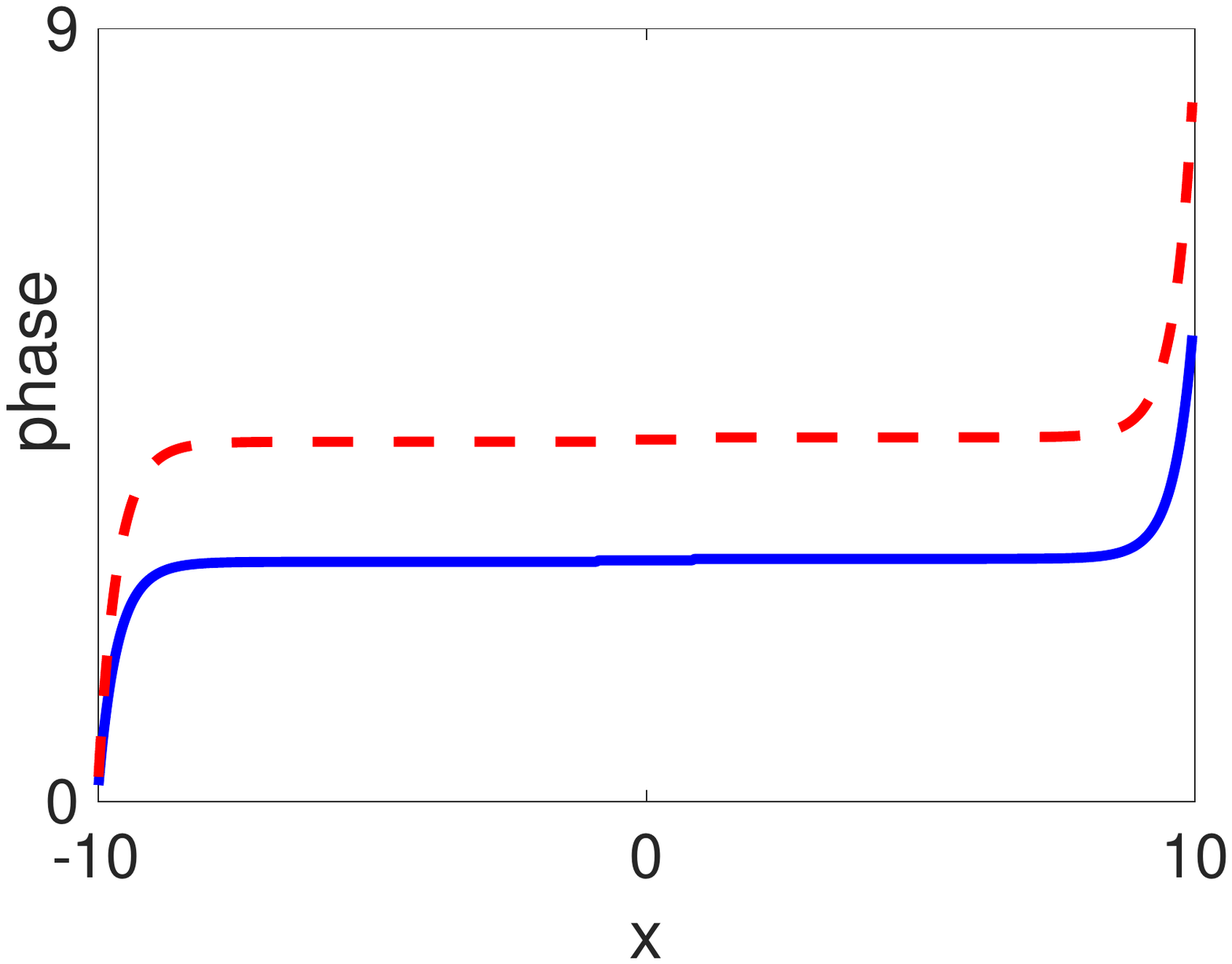}
\vspace{-3cm}
\caption{Multi-peak bright soliton solutions calculated for the parameters $h_{11}=2, h_{12}=1, h_{21}=1/2, h_{22}=2, W_1=-2, W_3=1.999999, \mu=-0.15, \sigma=-1$,  and $k=0.707$. Blue solid curve corresponds to the first component and red dashed curve corresponds to the second component. (Right) non-trivial phase $\theta_j$ (see Eq. \eqref{phase_multi_peak}). Blue solid curve corresponds to the first component and red dashed curve corresponds to the second component.}\label{fig_multi_peak}
\end{center}
\end{figure}

 Then, using \eqref{recti} and \eqref{sol_U} we find solutions of Eqs. \eqref{sys} of the form
\begin{equation}
\psi_j(t,x)=e^{\frac{\mu}{2} x^2}U_j(y(x))e^{i(\theta_j(x)+\mu t)}, \quad j=1,2.
\end{equation}
It is straightforward to show that $R_j(x)=e^{\frac{\mu}{2} x^2}U_j(y(x))\to 0$ as $x\to\pm\infty$ and that these are indeed localized solutions of our problem.

\begin{figure}
\begin{center}
\vspace{-4cm}
\includegraphics[height=13cm,width=6cm]{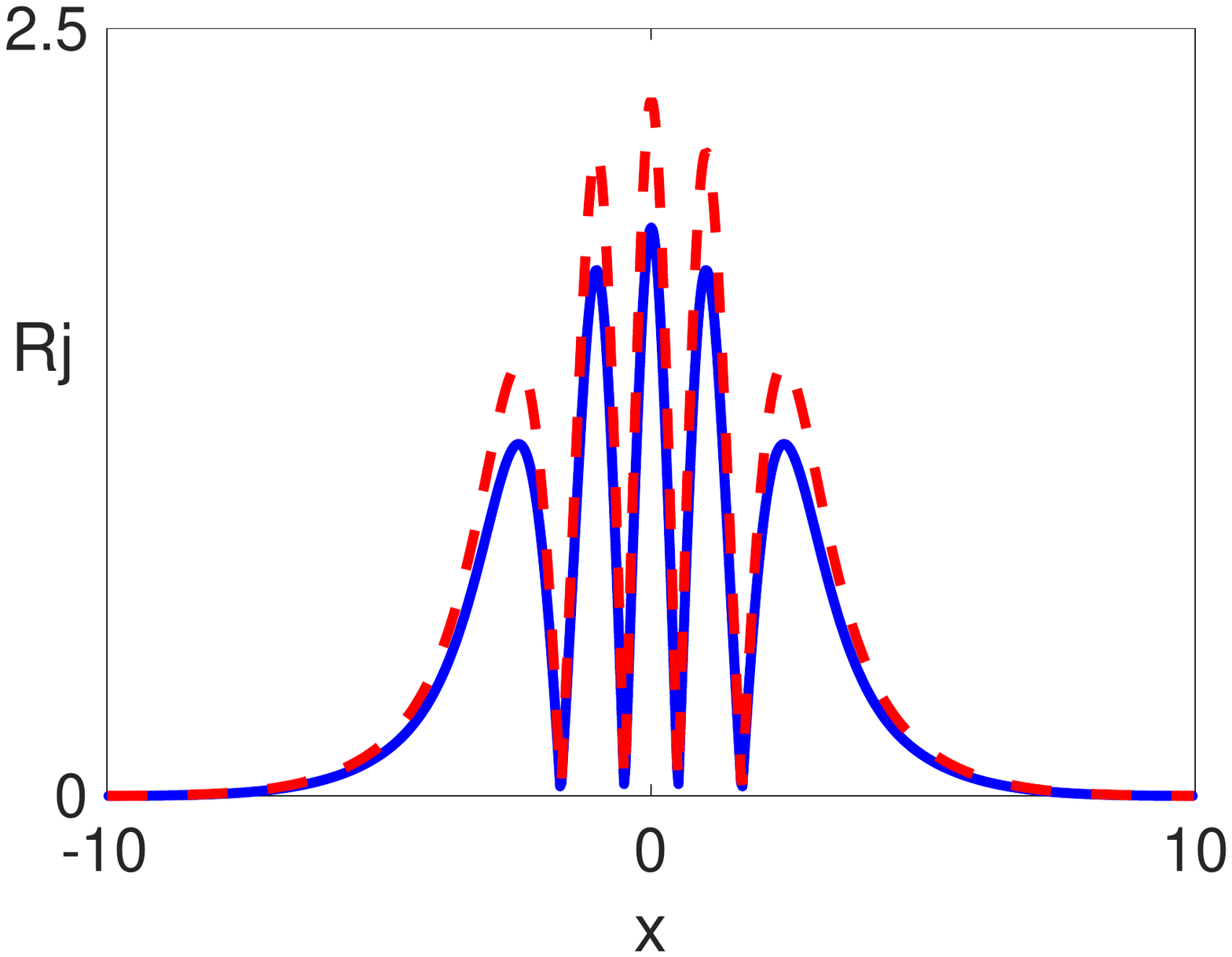}
\includegraphics[height=13cm,width=6cm]{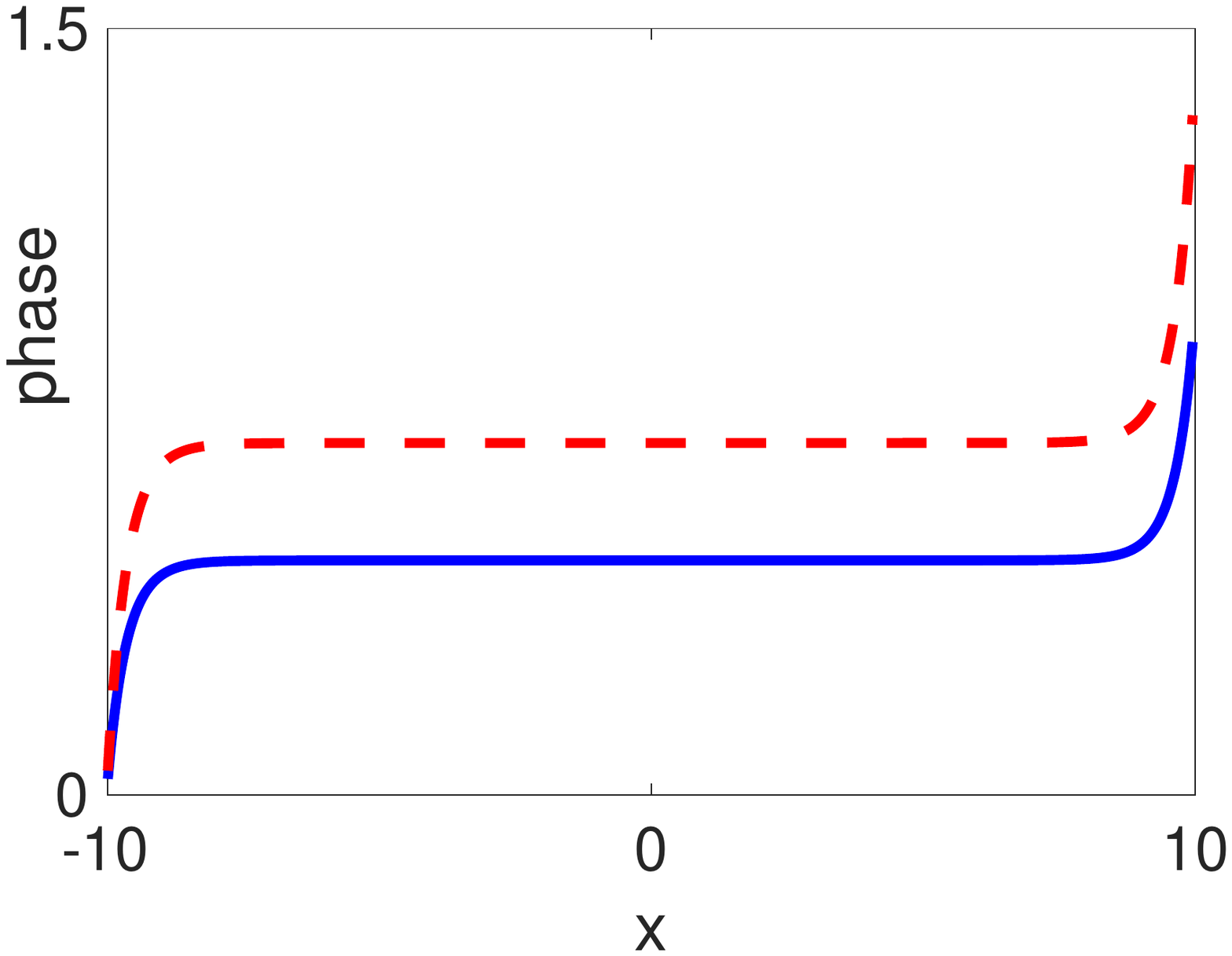}
\vspace{-3cm}
\caption{Multi-peak bright soliton solutions calculated for the parameters $h_{11}=2, h_{12}=1, h_{21}=1/2, h_{22}=2, W_1=-6, W_3=5.9999999, \mu=-0.15, \sigma=-1$,  and $k=0.707$. Blue solid curve corresponds to the first component and red dashed curve corresponds to the second component. (Right) non-trivial phase $\theta_j$ (see Eq. \eqref{phase_multi_peak}). Blue solid curve corresponds to the first component and red dashed curve corresponds to the second component.}\label{fig_multi_peak2}
\end{center}
\end{figure}

Fig. \ref{fig_bright_bright} (Left) shows a bright-bright soliton solution type and Fig. \ref{fig_bright_bright} (Right) its corresponding non-trivial phase $\theta_j$. With the parameters used in the solution (see caption of Figure \ref{fig_bright_bright}), we obtain, from Eqs. \eqref{constraints},
\begin{eqnarray*}
 W_2&=&-\left(W_1+W_3\right)=0.0499,\ C_0=-\left(W_1W_2+W_2W_3+W_1W_3\right)=0.0075,\\
  c&=&-W_1W_2W_3=2.49\ \cdot 10^{-4}.
 \end{eqnarray*}

 It is interesting to note that by varying the $W_i, i=1,2,3$ parameters, different bright-bright solitons solutions will be obtained, since the value of $k$ also changes. For example, Figure \ref{fig_multi_peak} (Left) shows the so-called multi-peak bright soliton solutions, which were calculated for $W_1=-2,\ W_2=1\cdot 10^{-7},\ W_3=1.999999$. The rest of the parameters were calculated in the same way as in the previous example.

 Also, Figure \ref{fig_multi_peak2} (Left) shows another multi-peak bright soliton solutions. Again, we would like to recall that these solutions present a non-trivial phase $(\theta\neq 0)$, which are different from other solutions calculated for trivial phase $(\theta\equiv 0)$ (Figs. \ref{fig_multi_peak} and \ref{fig_multi_peak2} (Right)).

\begin{figure}
\begin{center}
\vspace{-4cm}
\includegraphics[height=13cm,width=6cm]{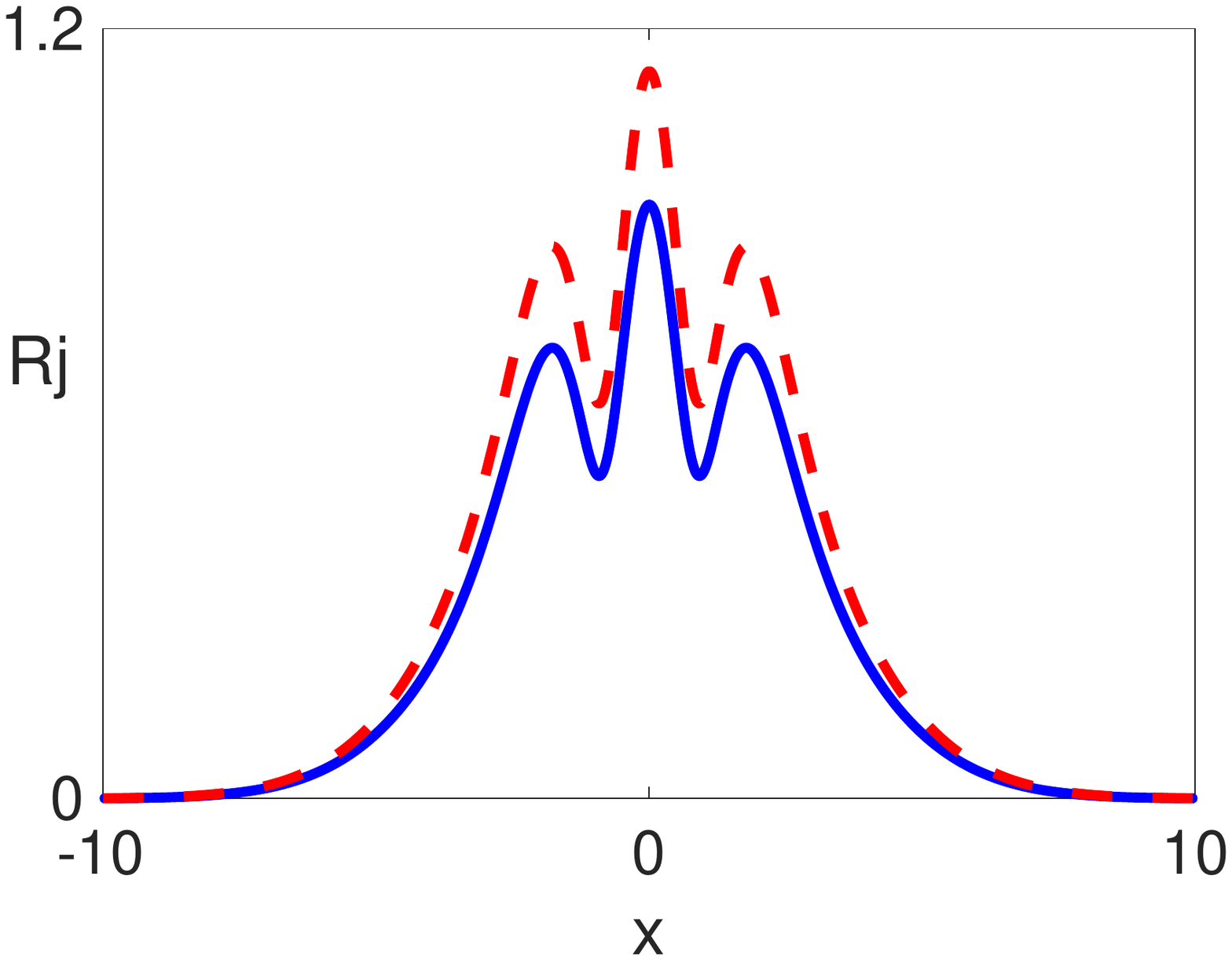}
\includegraphics[height=13cm,width=6cm]{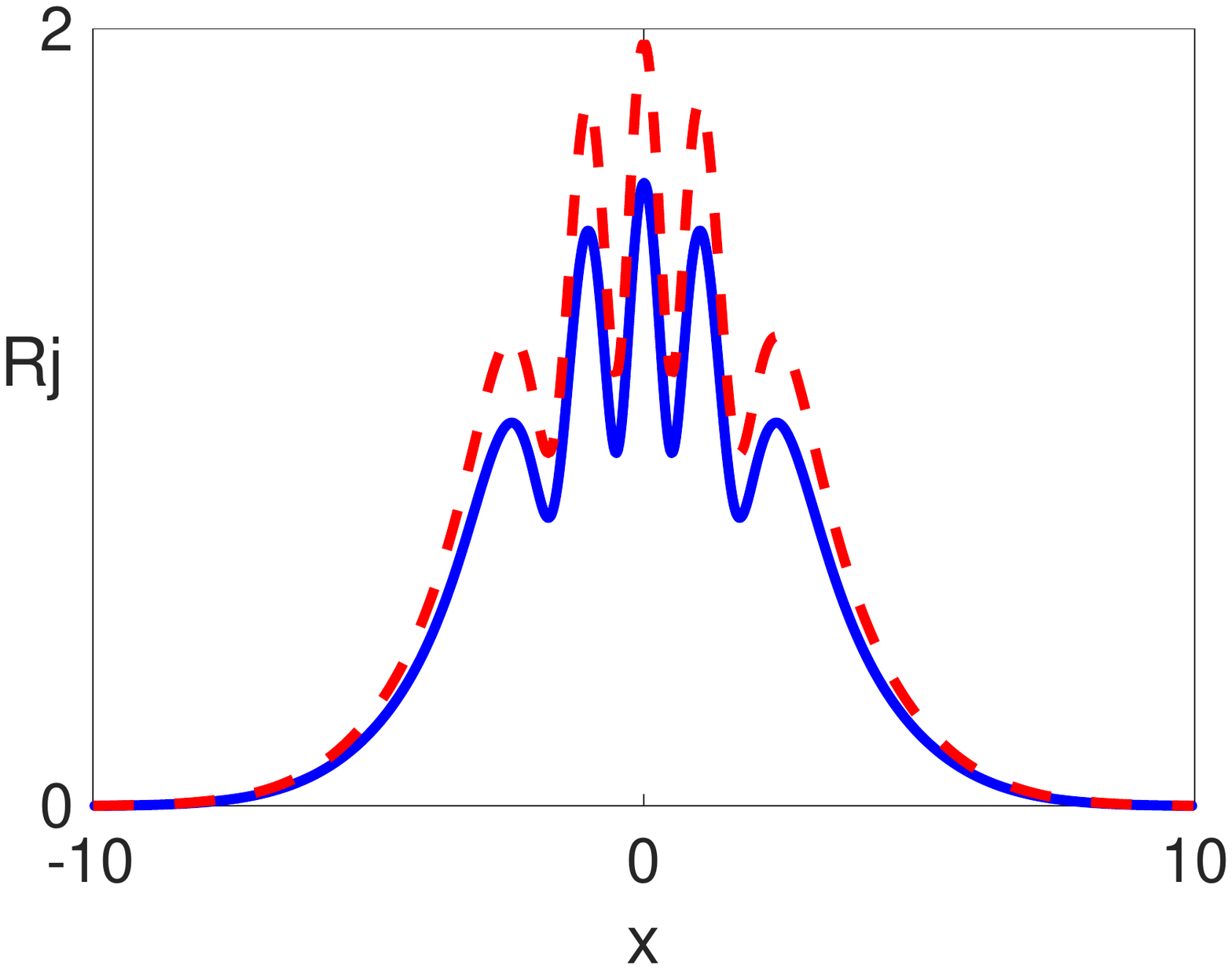}
\vspace{-3cm}
\caption{Multi-peak bright soliton solutions calculated for the parameters $h_{11}=2, h_{12}=1, h_{21}=1/2, h_{22}=2,  \sigma=-1, \mu=-0.15$ and (Left) $W_1=-2, W_3=1.5,$ and $k^*=0.534$ (Right) $W_1=-6, W_3=4.5,$ and $k=0.534$. Blue solid curve corresponds to the first component and red dashed curve corresponds to the second component.}\label{fig_multi_peak3}
\end{center}
\end{figure}

 Finally, in Figure \ref{fig_multi_peak3} we show other solutions which can considered as an ``intermediate solution'' between figures \ref{fig_bright_bright} and \ref{fig_multi_peak} (Left) and \ref{fig_multi_peak} and \ref{fig_multi_peak2}.

The non-trivial phase $\theta_j(x)$ was calculated using the formula
\begin{equation}\label{phase_multi_peak}
\theta_j(x)=\int^x \frac{c_j}{\delta_j^2e^{\mu s^2}\left(W_3-(W_3-W_2)\sn^2(\lambda y(s),k^*)\right)}ds.
\end{equation}


\section{Conclusions}

In this article, we have used Lie symmetries to construct explicit solutions with non-trivial phase of coupled non-linear Schr\"odinger systems with spatially inhomogeneous nonlinearities. First, the system of non-linear Schr\"odinger equations is reduced to an integrable singular ordinary differential equation. Once this equation is solved, we can explicitly calculate the solutions of the system of coupled Schr\"odinger equations using the solutions of the singular ODE.

Although we have restricted our attention to a few selected examples of physical relevance, the range of nonlinearities and potentials for which this can be done is very wide. We have constructed explicit solutions (dark soliton solutions) for $V=0$ with periodic nonlinearities. Moreover, we construct specific families of solutions (bright-- bright solitons and multi-peak bright soliton solutions) when the potential is quadratic on the space coordinates.

\section*{Acknowledgements}

This work has been partially supported by Spanish MICINN Grant with FEDER funds MTM2017-82348-C2-1-P. F. G\"ung\"or  expresses his gratitude to P.J. Torres for the hospitality and financial support that made possible a visit to the Department of Applied Mathematics, University of Granada, where the initial stages of this work were done.



\end{document}